# Project TAIPAN: Results from a Novel Gravity Gradiometer Field Test

**Alexey V. Veryaskin[1*], Howard C. Golden[1,2], Khyl J. McMahon[1,3], Neil M. Provins[1,4], Frank J. van Kann[2] and Thomas J. Meyer[5]**

[1] Trinity Research Lab, School of Physics, Mathematics and Computing, University of Western Australia, 35 Stirling Highway, Crawley, WA 6009, Australia

[2] Department of Physics, School of Physics, Mathematics and Computing, University of Western Australia, 35 Stirling Highway, Crawley, WA 6009, Australia

[3] AUX Pty Ltd, 65 Dundas Road, Inglewood, WA 6052, Australia

[4] MOROCK (WA) Pty Ltd, 2/47 Tully Road, East Perth 6004, Australia

[5] Lockheed Martin RMS – Gravity Systems, 2221 Niagara Falls Boulevard, Niagara Falls, NY 14304, USA

**Abstract**

Project TAIPAN has been carried out jointly by Trinity Research Lab and the Frequency and Quantum Metrology Research Group located at the School of Physics, Mathematics and Computing of the University of Western Australia (UWA). Lockheed Martin Corporation (USA) has also been a partner in this joint collaboration providing financial backing to the project and other support including advanced modelling, assessment of laboratory tests and data analysis. The project's aim was to develop a miniaturised gravity gradiometer to measure horizontal mixed gradient components of the Earth's gravity in a small, lightweight package that can be deployed in a fixed 4D mode, in a borehole, or on moving exploration platforms including ground-based, airborne and submersible. The gradiometer design has evolved through a few prototypes combining the design of its sensing element with ultra-low noise microwave and capacitive read-outs. The most recent prototype of the gradiometer using novel ultra-sensitive capacitive pick-off metrology has been trialled in the harsh environment of Outback Western Australia over a known gravity anomaly displaying steep gradients. Despite adverse weather conditions, results of the trial indicate that the gradiometer operated as expected, closely replicating the gravity gradient profile extrapolated from a regional gravity survey.

[*]Corresponding author: alexey.veryaskin@uwa.edu.au

## 1. Introduction

Gravity gradiometry has been an active multi-disciplinary area of research and development for more than 100 years that brings together pure and applied physics, precision engineering and ultra-low noise electronics. The best existing classic instruments have proved to be capable of providing a resolution of about a few Eotvos (1 Eotvos is the measure of gravity gradients equal to 1 x $10^{-9}$ 1/$sec^2$) in a highly turbulent environment, that is typical for airborne gravity gradiometry where the signal-to-noise ratio is a few parts in a billion [1]. Multiple critical applications of gravity gradiometry include natural resources exploration, oil reservoir monitoring and defence. The future trends in gradiometer sensor development include quantum-enabled systems, simplification, and miniaturisation which yields greater operational utility. The development of non-traditional gravity gradiometry technologies, whose processing or data interpretation may leverage artificial intelligence and machine learning, may also lead to significant reduction of the cost, weight, size and power consumption of the future instruments [2], [3].

Project TAIPAN was aimed primarily at bringing gravity gradiometry into the downhole environment, a goal that has been elusive due to the high temperature, high pressure, limited space and other difficult conditions in drilled wells. References exist in the open literature extolling the virtues of downhole gravity measurements for formation evaluation, deep reading of bulk density, profiling salt flanks, and monitoring fluid contact and movement. Ander and Biegert [4] report additional uses that could be added in the measurement-while-logging and measurement-while-drilling scenarios. In fact, reading these papers, one might conclude that gravity methods are ubiquitous in the well logging environment. However, nothing could be further from reality. As Chapin and Ander [5] pointed out nearly 20 years ago, gravity remains a fundamentally sound but underutilised technology. Black and Hare [6] describe the latest miniaturised scalar gravity sensor technology adapted for oil and gas use, citing that "gravity can now be measured practically anywhere a drill hole can go." The Micro-g LaCoste tool addresses the need for smaller size and greater range of well deviation (the tool can be used beyond horizontal). Improvements to measurement error stability and packaging for ruggedized use are also reported. Nonetheless, measurements must be made in a stop-and-dwell manner with precise positioning controls, so cannot be used as part of a wireline tool string nor included with a standard service. By and large many of the sensor technology gaps identified by Loermans and Kelder persist [7]. From these, one can deduce that it is a practical necessity for impactful industry uptake of the technology that any proposed gravity tool be compatible with standard well logging procedures. Considering the equivalence principle and lack of sufficient down hole inertial reference to distinguish gravity from tool motions (acceleration), it seems a heroic task to add scalar gravity sensing to wireline services. Enter gravity gradiometry for common mode rejection of unwanted accelerations and sharp focus on contacts and bulk density anomalies.

## 2. History

In 2017-2018 the first laboratory prototype of the proposed gravity gradiometer module was developed jointly by the Frequency and Quantum Metrology Research Group, Trinity Research Lab and Lockheed Martin Corporation which served as an industry partner to the University of Western Australia for about 6 years. The new TAIPAN gravity sensor is a non-articulated, monolithic ribbon-like sensing element with integral ultra-precision deformation gauging [8], [9]. It is built for simplicity and is amenable to the ruggedness needed for down hole use. The asymmetric response of the two half lengths of the sensing element is processed to give a gravity gradient while rejecting accelerations and uniform gravity input. Initially, a microwave interferometer, based on a pair of microwave re-entry cavities, was proposed to be used as the means of mechanical displacement sensing, providing a position noise floor of about $4\times10^{-14}$ m/√Hz. The first laboratory tests of the 2017-2018 prototype based on the microwave read-out successfully reproduced the forward modelled gravity gradient signals generated by a moving test mass in the laboratory environment [10]. However, as was soon realised, the microwave metrology used to convert the gradiometer's sensing element differential displacement into a microwave carrier phase modulation and then into digitised and recorded data sets would be difficult and costly to use for downhole applications.

In 2019-2023 Lockheed Martin and Trinity Research Lab were co-developing a new gravity gradient sensor that, along with deployment on small unmanned vehicles, could also find use as part of a standard wireline service to the oil and gas or minerals industry. Its slender form factor combined with an appropriately designed sonde interface provides the opportunity for true wireline use as part of a tool string, e.g. for open hole logging, free from the stop-and-dwell deployment forcing niche use of gravity sensors heretofore. Measurements are directional and encapsulate the change in gravitational attraction into the formation at a specific tool azimuth as it moves along the bore axis, i.e., an along bore-outward to formation gradient is available as a function of azimuth. The sensor's mechanical susceptibility is internally modulated giving spectral separation of the sought gravity signals from would-be low frequency errors owing to minor, but unavoidable, sensor bias and drift. The new self-contained advanced development gradiometer module was assembled in 2023 for field tests at ground surface, one step closer to the form factor needed for inclusion in a wireline tool string, usable at all bore deviations.

For the first time a novel capacitive sensing metrology has been used replacing the microwave cavity used in previous prototypes [11]. The innovation is that the new design has completely eliminated the parasitic AM-modulation caused by notch-shaped microwave or RF carrier frequency-to-amplitude conversion inherited in the older versions and responsible for the lack of long-term stability. The new capacitance-to-phase single chip converter is used as the capacitive sensor interface that contains only a single active component (a dual opamp) and a few passive components. This configuration provides a high steepness capacitance-to-phase conversion rate ( ~ 5 rad/pF ). Operating in a carrier frequency range between 1 – 3 MHz and fashioned as a complete mechanical displacement-to-voltage transducer with proper back-end signal processing avoiding the carrier phase noise, the corresponding mechanical displacement sensitivity is limited to a few femtometres/√Hz. All the electronic components of the novel capacitive read-out are also available for high temperature and high pressure applications and can be supplied off-the-shelf at a reasonably low cost. In reality, the thermally activated gravity gradient noise is the dominant sensitivity limiting factor for the developed TAIPAN gravity gradiometer module [9]. However, multiple simulations and modelling analysis for typical borehole applications have proved that a resolution of about 20 Eotvos at 1 sec average would be of a significant value for downhole applications and that is within the TAIPAN reach.

## 3. Preparation for and Conducting the Field Test

The field test was scheduled to be conducted between September 10 and September 14, 2023 at a site 622 km from Perth where public domain gravity survey data over an extended outback area were available beforehand. The data were assessed and an expected horizontal gravity gradient anomaly along the chosen test path was extrapolated from the 400 m x 800 m gravity survey grid stations (Fig. 1 and Fig. 2).

It is worth noting that the goal of the proposed field test was not to measure gravity gradients at the laboratory level of performance but rather to assess the sensor's tolerance to a realistic field environment. It was decided that the sensor will be deployed in a nearly strap-down operating mode and none of the normally used protective measures such as active angular stabilisation and vibration isolation should be used.

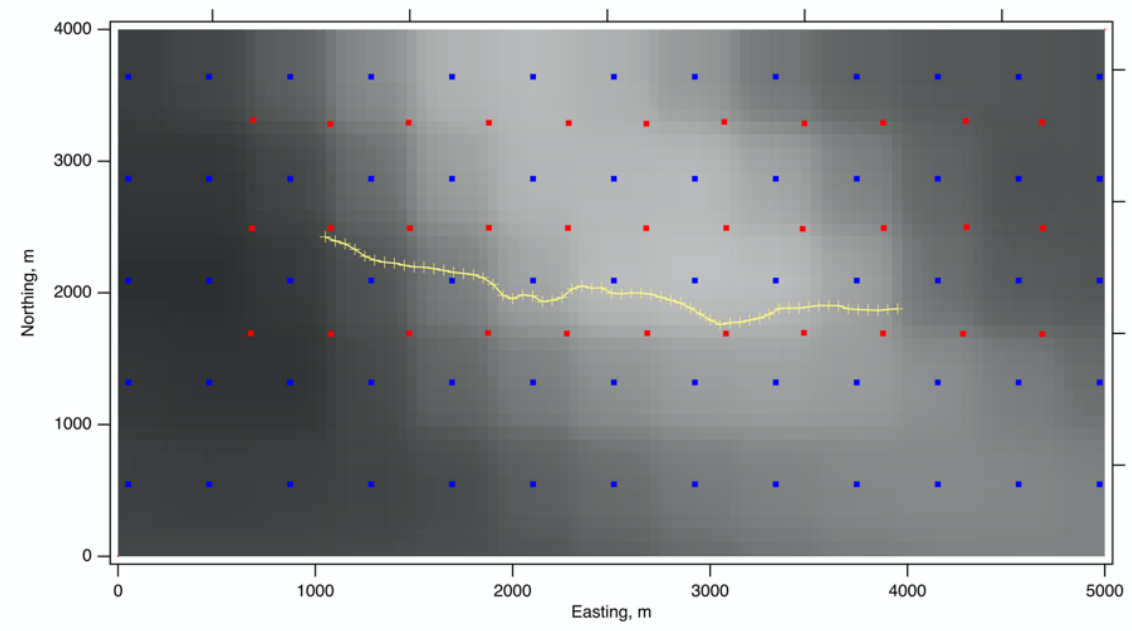

Figure 1. Red dots – original gravity stations, blue dots – interpolation grid points, yellow line – the measurement path, yellow crosses – the sample points.

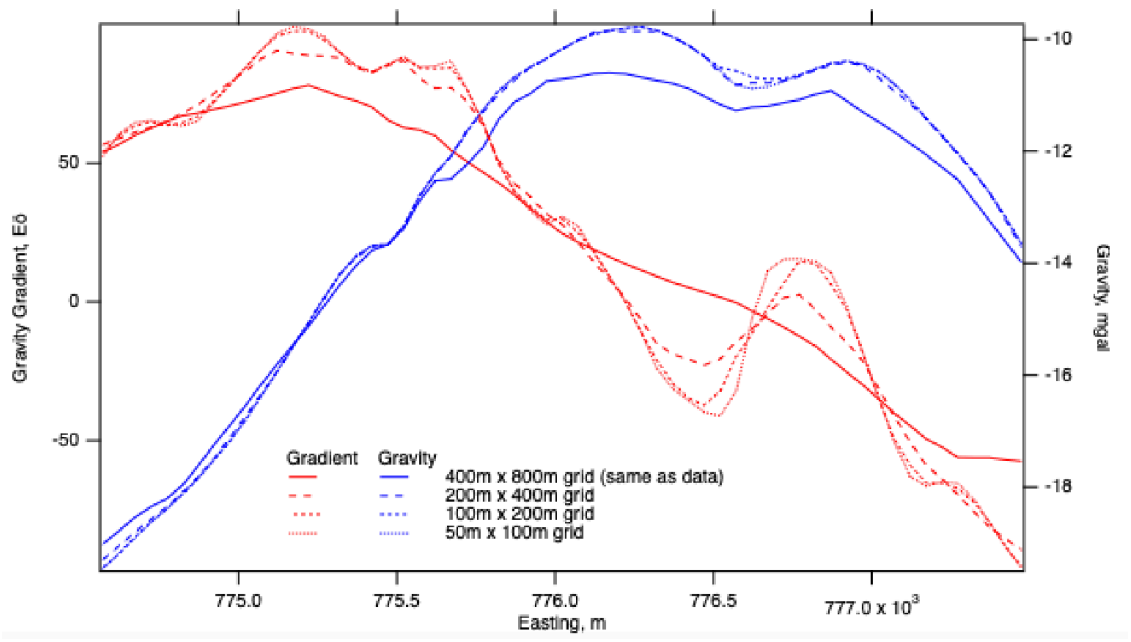

Figure 2. Extrapolation of the potential gravity gradient anomaly that would be expected along the test path profile based on the publicly available gravity survey data.

The sensor was carefully examined and a pre-field-test was conducted in an open space environment, detecting a medium size vehicle moving by the TAIPAN assembly at about 3 metres distance. This resulted in a clear gravity gradient signal detected by the TAIPAN signal processor (see the Supplementary Materials).

The complete TAIPAN sensor frame was mounted inside a commercial vacuum flask with sealed feedthroughs in order to isolate it from the outside contamination. After about 8 hours of driving the crew arrived at the test site on September 11, 2023. The TAIPAN sensor assembly was inspected and found to be functioning nominally, requiring no further adjustments to the TAIPAN core ribbon & frame assembly inside the vacuum flask. Only the vacuum was slightly degraded and was increased to a modest level of ~ 0.01 standard atmosphere.

The TAIPAN assembly was deployed the following morning, it was mounted upon a self-levelling trolley designed for and previously used in the harsh outback environment , as depicted in Fig. 3 and Fig.

4. Four air-pressure damping mounts were the only isolating components on the TAIPAN assembly.

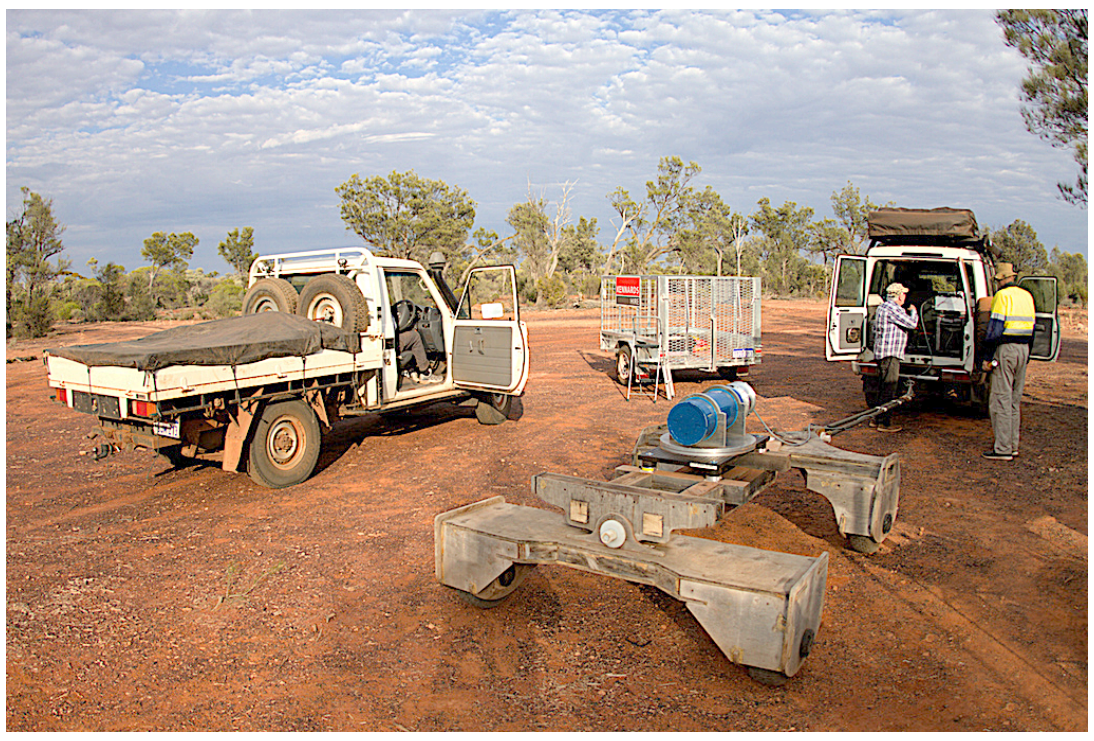

Figure 3. TAIPAN gradiometer set up prepared for the first run along the test path.

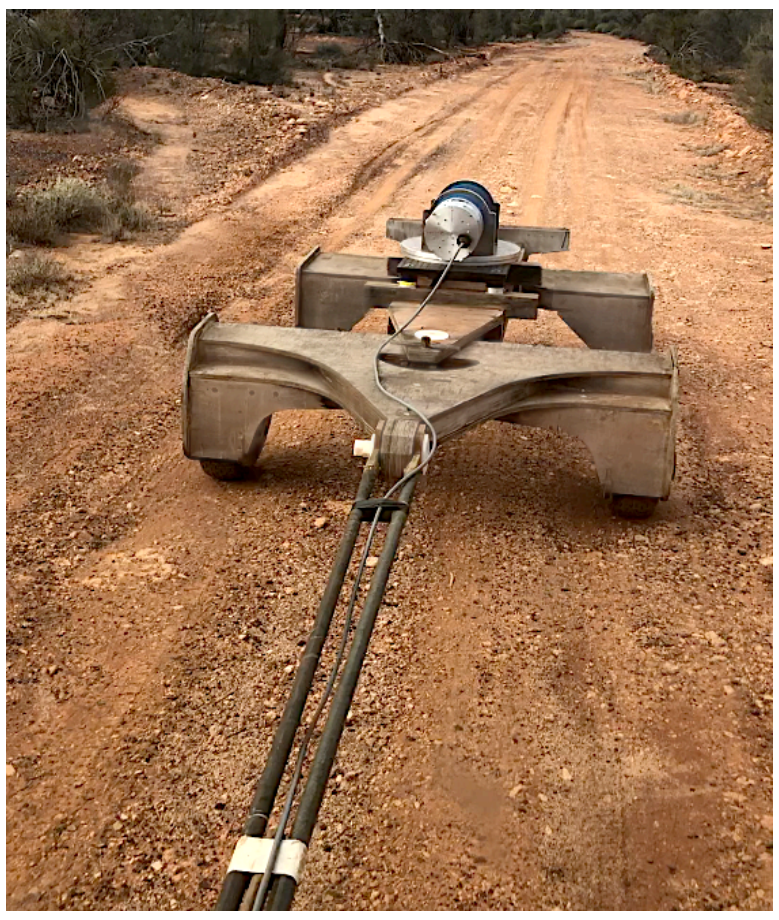

Figure 4. Moving to the first measurement station.

The TAIPAN operator and all back-end control and data logging equipment were placed in the towing vehicle behind which the trolley was towed. The TAIPAN sensor assembly contained a rotary table upon which the vacuum flask and the front-end electronics unit was strapped down. The measurement data were recorded over 2 sec measurement intervals only when the TAIPAN gradiometer set up was not moving, i.e. in a stop-and-dwell manner. The latter allowed alignment of the TAIPAN ribbon axis (say X axis) along a fixed GPS coordinate direction only.

A marine grade temperature sensor along with a solid MEMS dual-axis tilt meter were installed inside the vacuum flask in order to monitor and record the temperature of the TAIPAN sensor frame and the 2D tilting at each measurement station. In total, 60 stations separated by 50 metres distance were measured along the test path of about 3 km long as depicted in Fig. 5 below.

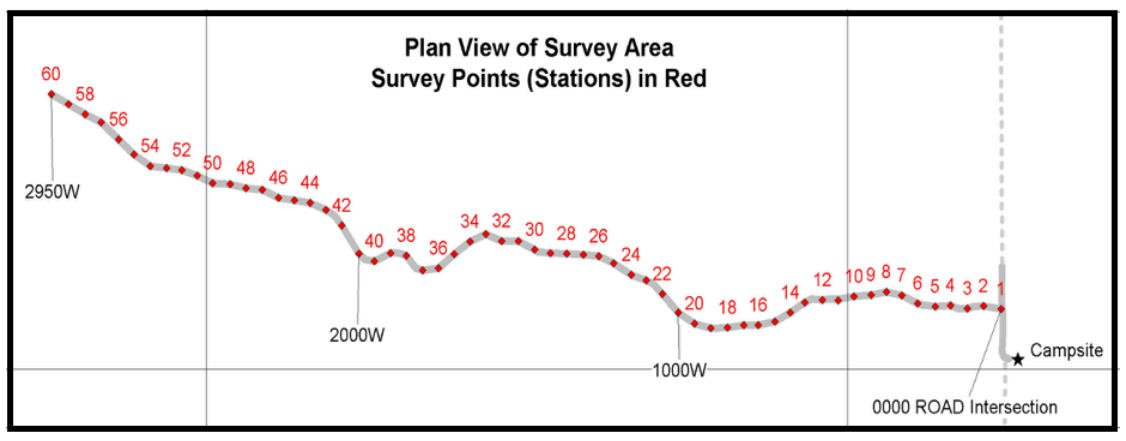

Figure 5. 60 measurement stations separated by 50 metres gap have been marked-up along the test path.

On September 12, 2023 (early morning) the measurements at the 60 stations for both the east-west profile or the west-east profile (return) were completed. This took about 6 hours of continuous stop-and-take-a-measurement operating time for the TAIPAN sensor assembly. At each station the orientation of the sensor longitudinal axis (X-axis) was aligned with the East-West or West-East directions. The return orientation was 180 degree shifted in the horizontal plane compared to the forward one.

## 4. TAIPAN Electronics and Data Processing Set Up

The complete block diagram of the front-end and the back-end of the signal processing chain is shown in Fig. 6 below. The TAIPAN operator manually controlled back-end equipment comprised of a 500 Watt 12 V power station, a tough touch screen, a PC controller (Linux), a PicoScope5000 data logger & oscilloscope and a MOKU.GO (Liquid Instruments, USA) Frequency Response Analyser with a built-in PLL lock-in amplifier unit.

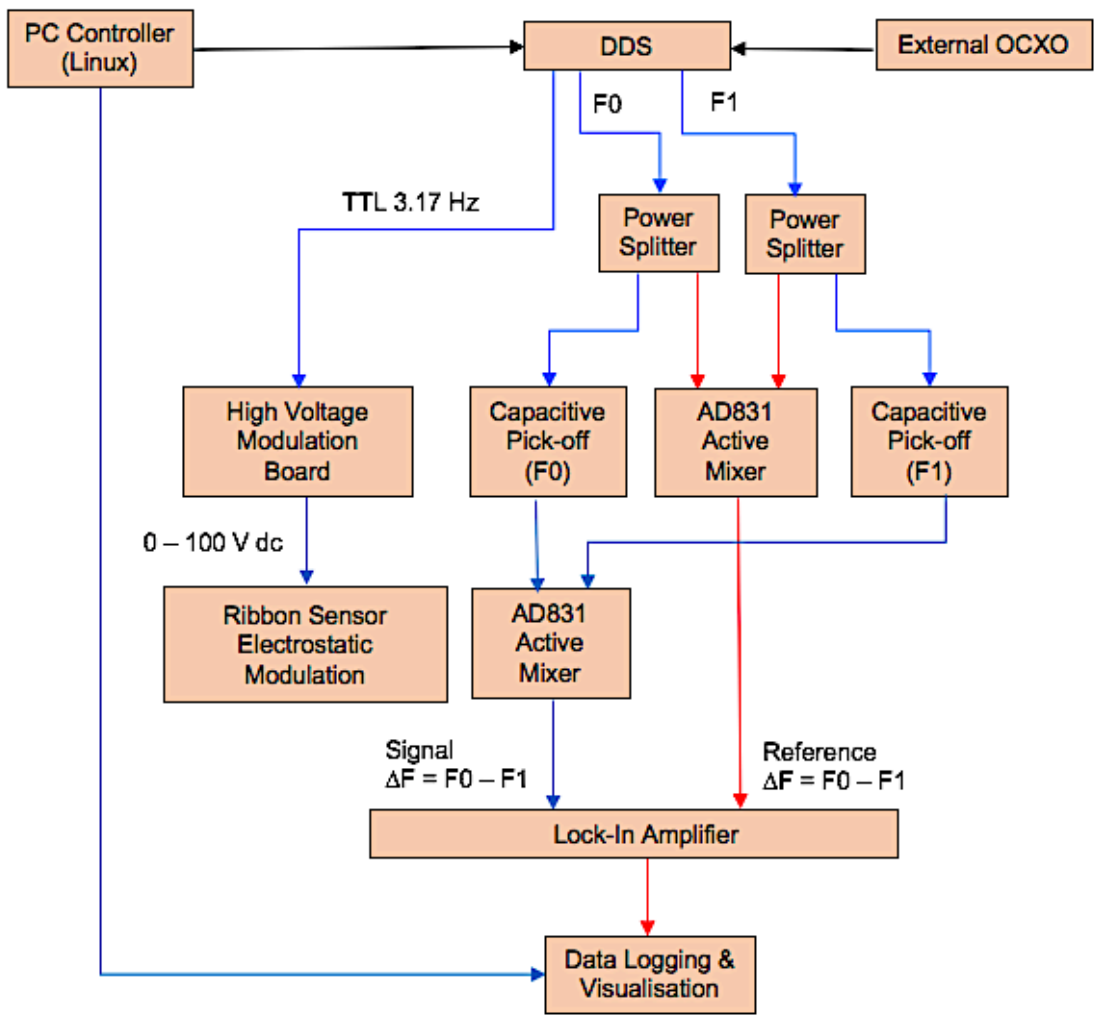

Figure 6. The complete block-diagram of the TAIPAN signal processing chain.

A frequency response analyser was used as a compact field deployable test unit in order to set *in situ* the correct RF carrier frequencies F0 and F1 tuned to the maximum ribbon ends displacement-to-phase conversion rates. A compact 4-channel (synchronised) direct digital synthesiser (DDS, Highland Technology, USA) was used in the field test providing the RF carrier signals to two individual capacitive pick-off elements coupled to the ends of the ribbon sensing element as shown in Fig. 7.

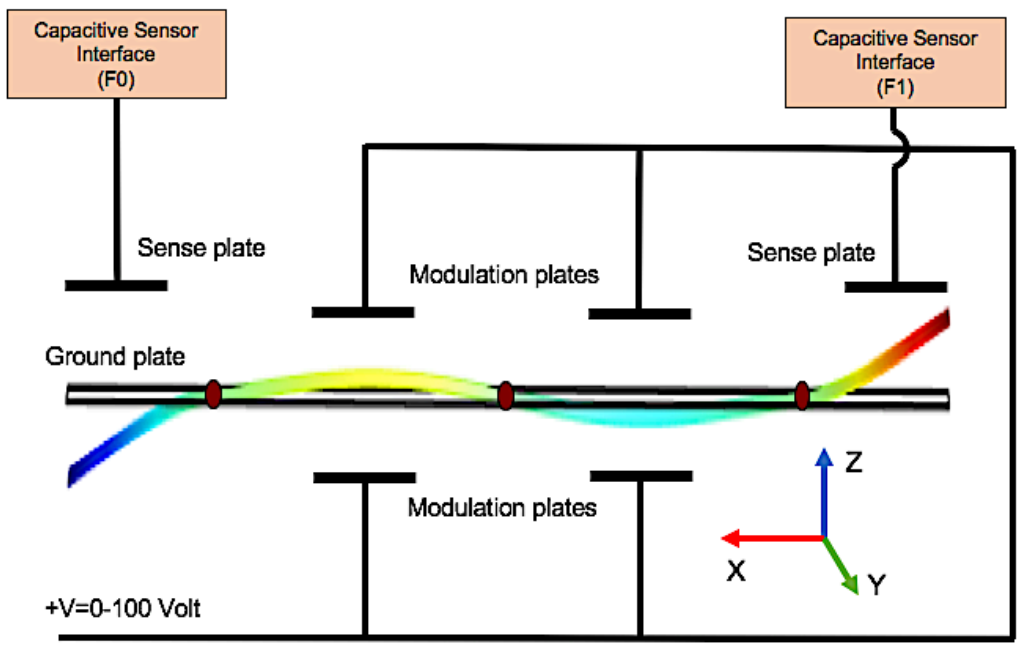

Figure 7. The TAIPAN sensing element (ribbon) is coupled to sensing and modulating capacitive plates. The shown free-hinged-hinged-hinged-free deformation profile is the result of a linearly distributed force-per-unit-length, representing a pure gravity gradient load.

A heterodyne detection approach was used allowing for the down conversion from the individual RF frequencies F0 and F1 to a differential frequency F0-F1 (typically within 5 – 10 kHz range). The latter is the same for both the reference phase-fixed signal and the phase-modulated one the differential phase of which is proportional to the synchronous displacements of the ribbon's ends. Applying the reference and phase-modulated signals to the final lock-in stage of the signal processing allowed for the DDS phase noise suppression in the demodulated differential phase component at the lock-in output. The latter was then digitised in a 16-bit data logger and visualised by using the PicoScope5000.

The TAIPAN signal processing was accomplished with a stiff-to-soft modulation technique provided by a set of horizontal capacitive plates positioned symmetrically with respect to the sensing element (ribbon) midpoint as depicted in Fig. 7. A TTL 3.17 Hz square wave signal was taken from the DDS and a high voltage drive module was used to convert positive ~ 2.8 V voltage level into ~ 100 V pulse that results in a negative spring effect applied to the ribbon's mechanical stiffness. The stiff and soft modulation cycles cause well-defined ring-down response of the TAIPAN ribbon sensor within the equal durations of cycles. The sensor becomes more sensitive to external forces in its soft stage and less sensitive otherwise. The demodulation process is a curve-fit algorithm that removes the ribbon's ring-down response from the data and then subtracts the averages of each cycle signal level one from the other. The result is proportional to the total differential force acting on the ribbon within the total (on plus off) duration of the TTL cycles. This modulation-demodulation algorithm was specifically designed to cancel out 1/f noise and/or zero-point drift in recorded data sets. It was successfully used for the first time in the laboratory environment in 2009 - 2010 by Gravitec Instruments Pty Ltd (Australia) and QinetiQ Ltd (UK) working jointly on a project aimed at the development of borehole gravity gradiometer based on the same principle as the TAIPAN sensor module [12]. However, the field test data collected in a harsh and unstable open space environment turned out to be much noisier compared to that collected in the laboratory. The curve-fit demodulation algorithm has been modified and adapted to an average-fit one applied to every modulation on-off cycle where the ring-down response is not well defined due to noisy conditions. A detailed description of the demodulation process and post-processing the recorded data sets is presented in Supplementary Materials to this paper.

The total differential force acting on the TAIPAN's sensing element consists of external differential force (gravity gradient-related force) and one caused by the high voltage applied to the modulation capacitive plates due to their residual mismatch. The total acquired signal can be represented as follows:

<u>East to West measurement data:</u>

$$V_{i,forward} = K_{total}(\Gamma_{zx}^{0} + \Delta\Gamma_{zx} + \delta\Gamma_{zx}) + Noise$$

<u>West to East measurement data:</u>

$$V_{i,return} = K_{total}(-\Gamma_{zx}^{0} - \Delta\Gamma_{zx} + \delta\Gamma_{zx}) + Noise$$

where $Vi$ is the averaged output signal at each station,

$Ktotal$ is the gradiometer's total gain, $\Gamma_{zx}^{0}$ is the local Earth gravity gradient component, $\Delta\Gamma_{zx}^{0}$ is the gravity gradient anomaly at each station and $\delta\Gamma_{zx}$ is the dynamic offset gravity gradient due to the electrostatic modulation mismatch errors.

The noise constituent in the recorded data has different components ranging from the fundamental thermally activated motion noise of the sensing element and noise in signal processing electronics to environmental noise factors such as windy conditions, tilting and heating.

The unknown offset gradient related part of the recorded signal can be removed to some level by taking the difference between the two data sets above, i.e.

$$V_{i,forward} - V_{i,return} = 2K_{total}(\Gamma_{zx}^{0} + \Delta\Gamma_{zx}) + Noise$$

This results in the total signal data proportional to the total gravity gradient along the measurement profile. The latter also assumes that the gradiometer's total gain and the dynamic offset gradient are both constant over the total measurement time of ~ 6 hours. This may not hold as there was a temperature drift inside the vacuum flask caused by the changing temperature outside (from 19.6 deg C to 37.5 deg C over 6 hours). Despite this significant temperature change the TAIPAN sensor was still functioning within its linear dynamic range. The total gain can be treated as a constant factor due to the stiff-to-soft modulation in place. However, this statement requires further testing with much longer operating time.

After applying the modified demodulation algorithm to the recoded rough field data, they were smoothed down by a native Igor Pro built-in binomial-fit procedure that removes short-term variations, or "noise" to reveal the important underlying unadulterated form of the data. The smoothed return data set was subtracted from the smoothed forward data set as shown above and the result comparing the predicated gravity gradient profile and the measured one is shown in Fig. 8 below

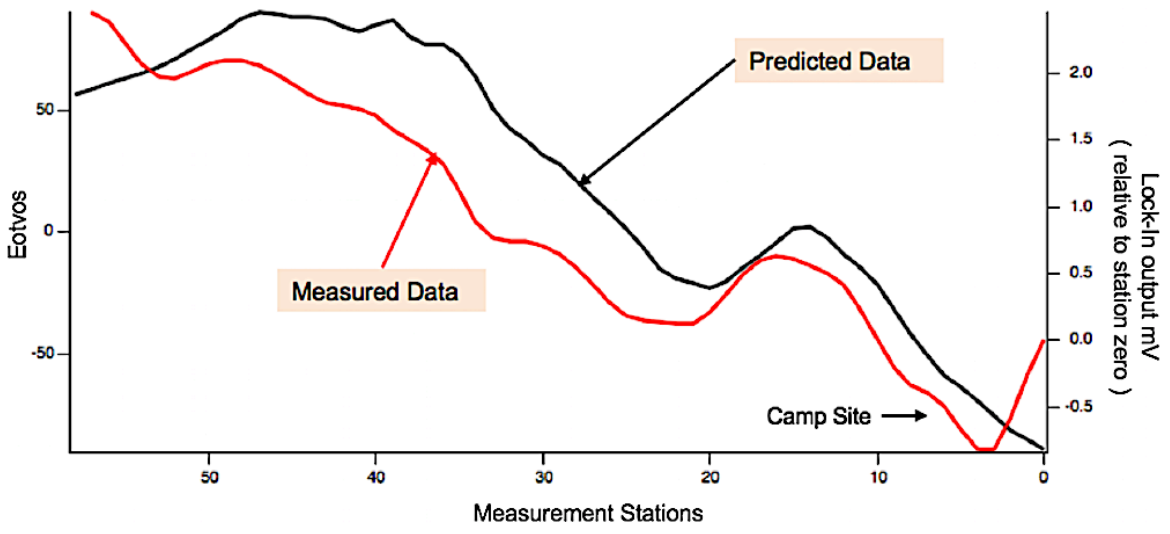

Figure 8. The smoothed gravity gradient profile versus the predicted one using the extrapolated gravity data. The gap between the two curves can be a result of the GPS uncertainty of the measured and predicted data sets.

Both the measured and extrapolated gravity gradient data depicted in Fig. 8 above are not well quantified and only demonstrate a similarity in their profiles along the test path. The predicted variations in the gravity gradient along the latter are within 50 – 200 Eotvos which are above the TAIPAN laboratory real time noise level (see Supplementary Materials). The real time data sets collected during the field test were much noisier and a maximum possible noise reduction treatment has been used to obtain the gravity gradient profile shown in Fig. 8.

## 5. Discussions

There are a number of error factors that appear in the cumulative data set resulting from the subtraction of the forward and return data recorded at different times and under different conditions. The total error budget analysis related to the field test data is not possible at this development stage as it lacks the definitive knowledge of all possible error sources affecting the performance of the TAIPAN gravity gradiometer module during the field trial. Nevertheless, most significant error factors are discussed below.

Different horizontal orientations and tilting of the TAIPAN sensor assembly result in different projections of the local gravity load upon the ribbon sensing element at the measurement stations. In turn, this results in pseudo-gradient response due to a limited Common Mode Rejection (CMR) of the TAIPAN sensor. The TAIPAN assembly horizontal position at each measurement station was assessed by measuring the angle between the local plumb line and the real time sensitivity axis of the ribbon sensor ( x-axis, see Fig. 7 ). A high resolution solid MEMS inclinometer DAS-30 ( Level Developments Ltd, UK ) was used to measure the tilting related variations of the normal to the ribbon sensing element component of the local acceleration of gravity. The corresponding error in assessing the measured horizontal gravity gradient related data then can be calculated as shown below [9]

$$\delta\Gamma_{zx} = \frac{\Delta g}{L}\ 10\text{^}\left(\frac{CMR}{20}\right)$$

where Δg is the normal to the ribbon projection of the local gravitational acceleration, L is the ribbon's length and CMR is the TAIPAN gradiometer common mode rejection in dB.

The maximum Δg variation along the test path was found to be ~0.8 mg ( see the Supplementary Materials ). Assuming the TAIPAN gradiometer CMR factor is minimum -120 dB, as this was measured in the laboratory environment, and using the current ribbon sensor length ( 0.3 m ), one can estimate the error associated with non-uniform gradiometer orientation along the test path, which is ~ 23 Eotvos.

Another error comes from the limitations in extrapolation of the potential gravity gradient anomaly that would be expected along the test path profile based on the publicly available gravity survey data. Assuming that the gravity survey was performed to currently available standards of accuracy, the error is likely to be approximately 0.1 mgal rms. When interpolated to a 50 m grid, this corresponds to ~20 Eotvos uncertainty. This is consistent with Fig. 2 which shows how the estimate of the gravity gradient converges to a single curve as the grid is made finer. The change in gradient when reducing the grid spacing from the original 400 m sampling grid to the finer one of 200 m is ~ 20 Eotvos.

## 6. Summary

Field testing was successfully carried out on a lightweight compact gravity gradiometer based on the deformation of a ribbon-like mass in the presence of a gravity gradient, the displacement of which was detected using capacitive metrology. The test was carried out in Western Australia over a known gravity anomaly with a steep horizontal gradient

As has been noted above, the goal of the field test was not specifically to measure the local gravity gradient profile with a good signal-to-noise ratio but rather to examine how different components of the TAIPAN sensor would function in a harsh unstable environment over a long operating time.

Taking advantage of a known gravity anomaly, this test proved appropriate for the goal. The sensor survived transport and operation in a real-world environment and, importantly, measured gravity gradients that correspond to modelled gradients based on actual gravity data. Although expected drift was encountered in the rough field data, this test was undertaken with no vibration, tilting or temperature stabilisation resulting in the TAIPAN gravity gradiometer performance that exceeded expectations.

Important challenges going forward include continued testing to ensure long-term repeatable results in a variety of environments, and packaging the gravity gradient sensor in a small, robust configuration enabling testing in realistic borehole conditions*.

## Acknowledgements

Author Veryaskin and author Golden are thankful to Dan Di Francesco of Niagara Gravity Consulting LLC for reading the manuscript and providing useful comments and suggestions.

## Data Availability Statement

The data cannot be made publicly available upon publication because they are not available in a format that is sufficiently accessible or reusable by other researchers. The data that support the results of the field test are available upon reasonable request from the authors.

## Authors Contribution

The first four authors comprised the crew that conducted the field test described here. Author Veryaskin was responsible for most of the technical content of the manuscript. Author van Kann analysed the available gravity data at the test site and extrapolated them to the expected gravity gradient anomaly along the test profile. He independently reviewed the recorded data sets and provided data analysis. Author Meyer wrote the introduction to this paper. All authors contributed significantly to the final content of the manuscript, providing critical input and amendments.

*Currently this development is put on hold due to permanent relocation of the Trinity Research Lab to New Zealand where it will be acting under a new business name GRADIOLAB.

## SUPPLEMENTARY MATERIALS

### THE PHYSICAL MODEL

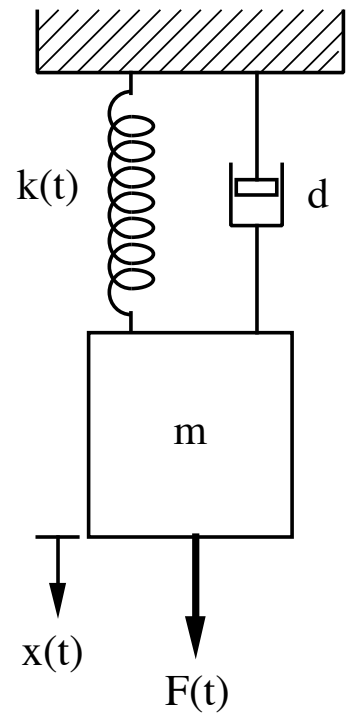

Figure 1. A simplified model of the physical system, depicting a single normal mode

The aim is to determine the force $F(t)$ from a measurement of $x(t)$ using a resonant capacitive transducer. The spring constant is modulated by the voltage $V_B(t)$ applied to the capacitive transducer

$$k(t) = k_0 - a\, V_B^2(t) \quad (1)$$

where $k_0$ is the natural spring constant and a describes the strength of the transducer coupling to the modulation process. The modulation is produced by a square wave at about 3 Hz, which is passed through a low pass filter before amplification to high voltage and the amplified signal is applied to the transducer. The frequency response of the amplifier coupled to the capacitive is unknown and must be determined either experimentally or by modelling the amplifier. It is possible that the damping is also modulated, but this cannot be confirmed until some preliminary processing has been completed.

The displacement $x(t)$ is sensed by a resonant capacitive transducer and demodulated to produce the measured signal $V_x(t)$. The demodulation produces an unknown delay, which must be determined either experimentally or by modelling the transducer and demodulation system.
The differential equation is

$$m\frac{d^2}{dt^2}x(t) = F(t) - k(t)\,x(t) - d\frac{d}{dt}x(t) \quad (2)$$

and this is non-linear, requiring a numerical solution.

The force $F(t)$ is slowly varying and can be assumed constant during each period of the bias voltage.

### THE DATA

Fig. 2 shows an extract of the acquired data under stationary condition before the field trial commenced. The grey trace represents the input to the HV amplifier, and the red trace represents the demodulated data. The reason for the apparent delay is unknown but the delay must be included in the model.

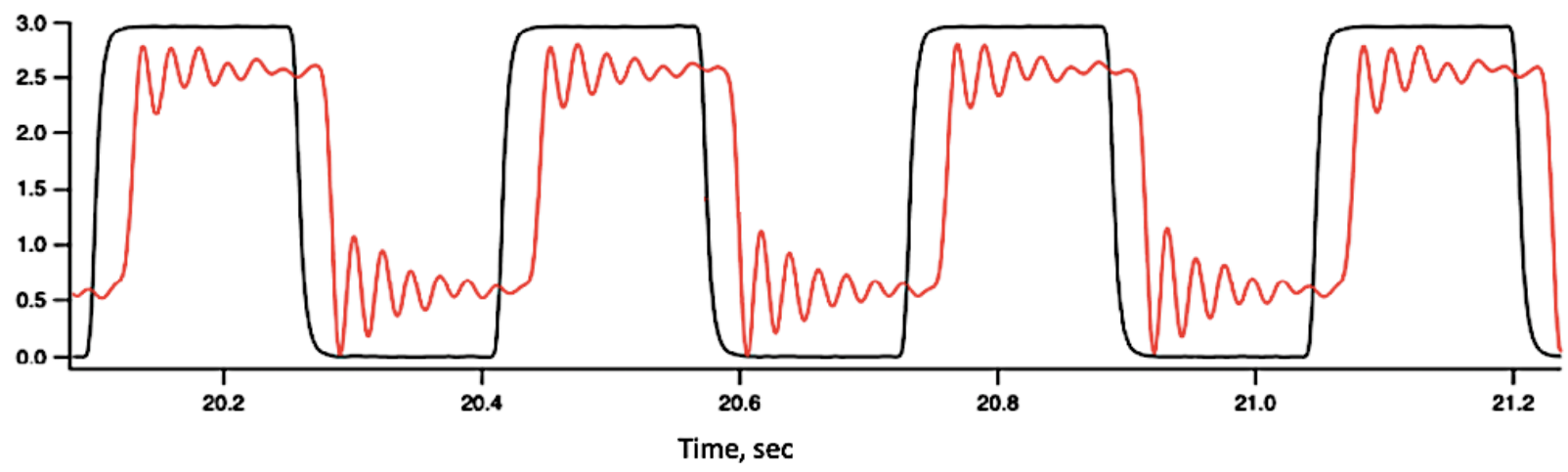

Figure 2. Extract from the pre-field-test data recoded under stationary conditions

THE MODEL

The output signal $V(t)$ is assumed to be proportional to the displacement delayed by a fixed latency $\Delta t$

$$V(t) = \sigma\, x(t - \Delta t) \qquad (3)$$

This assumes that the latency is due to the demodulation electronics and not the modulation of the stiffness. Equation 2 can be recast in the form

$$\frac{d^2}{dt^2} x(t) = A(t) - f_o^2\left(1 - \alpha\, V_B^2(t)\right) x(t) - \beta \frac{d}{dt} x(t) \qquad (4)$$

Defining $y(t) = \frac{d}{dt}x(t)$,, this gives two coupled first order differential equations

$$y(t) = \frac{d}{dt} x(t)$$
$$\frac{d}{dt} y(t) = A(t) - f_o^2\left(1 - \alpha\, V_B^2(t)\right) x(t) - \beta\, y(t) \qquad (5)$$

Equation 5 contains several free parameters, including α and β, which are simple scalars and A(t), which can be describes in terms of a list of discrete values

$$A_i = A\left(mod\left(t_j, T\right)\right)$$

where T is a characteristic time interval, during which A remains constant, and can be taken to be an integer multiple of the period of the modulating square wave. The index $j$ represents the discrete samples of the continuous signal in equation 5, such that

$$t_j = j\delta t,\ x_j = x\left(t_j\right), y_j = y\left(t_j\right)$$

$$and\ V_{B,j} = V_B\left(t_j\right),$$

where δt is the interval between samples.

The differential equation 5 can be solved if the parameters $f_o$, α, β and $A_i$ are known and the predicted output signal can then be obtained from equation 3, which introduces two initial parameters Δ and σ. This predicted signal can be compared with the measured output signal and the parameters can then be adjusted using a least-squares algorithm to make the predicted signal match the observed signal. This procedure is computationally expensive, as the full non-linear differential equation must be solved for each iteration of the fit. The parameter σ cannot be obtained independently from the data and must be determined by calibration. The rapid change in equilibrium position of the mass due to the rapid change in $k(t)$ excites the resonance, which rings down with a time constant longer than the half-period of the square wave, the processing is required to extract the equilibrium position from this ringdown.

THE SIMPLIFIED MODEL

Figure 2 indicates that the frequency of the resonant mode, which is excited by the modulation is the same for the rising edge and the falling edge, which is inconsistent with the equation 1, which predicts that the fundamental frequency for the two cases is different. In practice, there are numerous resonant modes in the system and perhaps the fundamental mode predicted by $k$ and $m$ is not the dominant one. Indeed, Fig. 1 depicts a lumped system, whereas in practice it is distributed, giving rise to more modes than the system depicted there, not all of which are modulated by the capacitive transducer. Therefore, it might be unnecessary to solve equation 5 for the two resonant frequencies corresponding to the two modulation states, and the observed resonant frequency can be obtained from the data, which is sufficient to determine the equilibrium position, by performing a curve fit to the ringdown, assuming that

$$x(t) = A\, e^{t/\tau}\ sin(2\pi f_o t + \varphi) \qquad (6)$$

with $A$, $f_o$, τ and φ determined from the data for each step.

The change in the value of $A$ before and after each step depends on two factors. The first is the value of the external force being applied (ideally through the gravity gradient) and the second is forces applied by the modulation capacitor, due to imperfect geometry. Ideally, the force exerted by the capacitor on the ribbon is zero but there will be a small residual force due to the geometrical imperfections. However, this is unimportant provided that this force remains constant for each step. The algorithm for extracting the equilibrium positions is described in the next section.

## THE PROCESSING ALGORITHM

The data may contain many steps and the main requirement is to identify these steps and process each one individually. An example data set is shown in figure 3.

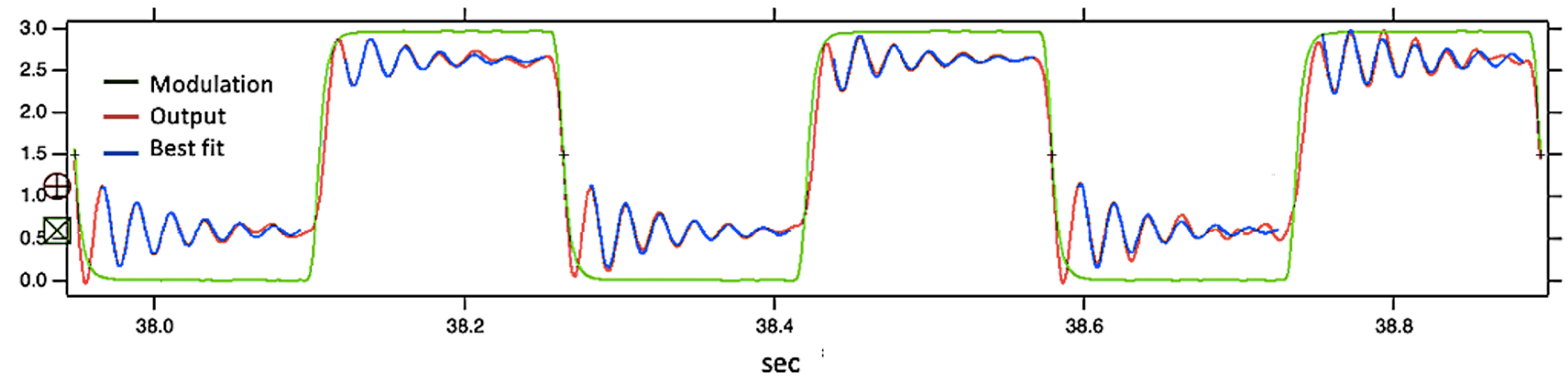

Figure 3. An extract from the processing record

In this figure, the modulation and output signals have been synchronised. This permits easy identification of the corresponding step in the output data. For each step, the equilibrium position is found using a least square fit to a subset of the response data. The first period of the ring-down is ignored, because it is affected by the smoothing of the modulation signal. The non-linear least-squares fit to a sinusoid is not straight forward because if the starting values of the parameters are not sufficiently close to the correct ones, then the fit may not converge and return incorrect parameters.

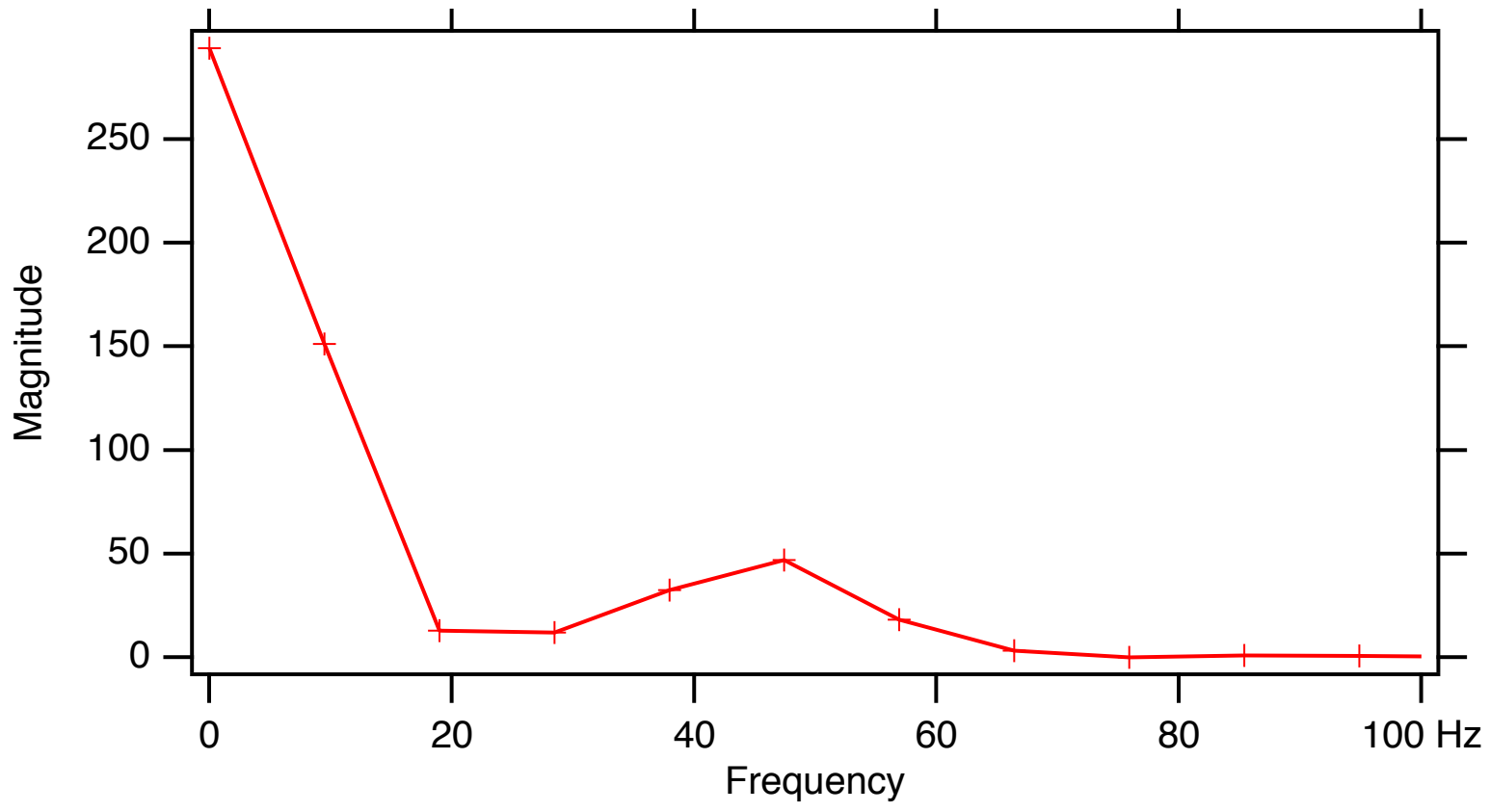

Figure 4. A typical Fourier transform

A Fourier transform of the subset of data is used to obtain the first estimate of the parameters and the magnitude of a typical transform of one of the steps is shown in Fig. 4, indicating a resonance at about 45 Hz. However, because the record is short, containing only a few cycles, the estimated frequency might not be sufficiently close for reliable curve fitting. Therefore, a simplified fit is performed, which ignores the ringdown and uses a simple sinusoid to obtain a better estimate of the frequency.

Finally, a full fit is performed to obtain all four parameters in equation 6 and this converged successfully for each of the 300 steps in the data record, and these parameters are stored as a time sequence. The desired parameter is the equilibrium position following each step and the corresponding sequences for these are shown in Fig. 5. The green crosses show the position corresponding to the off state of the modulation, while the blue crosses show that for the on state. The corresponding size of the step is shown as the red curve and should be proportional to the applied force.

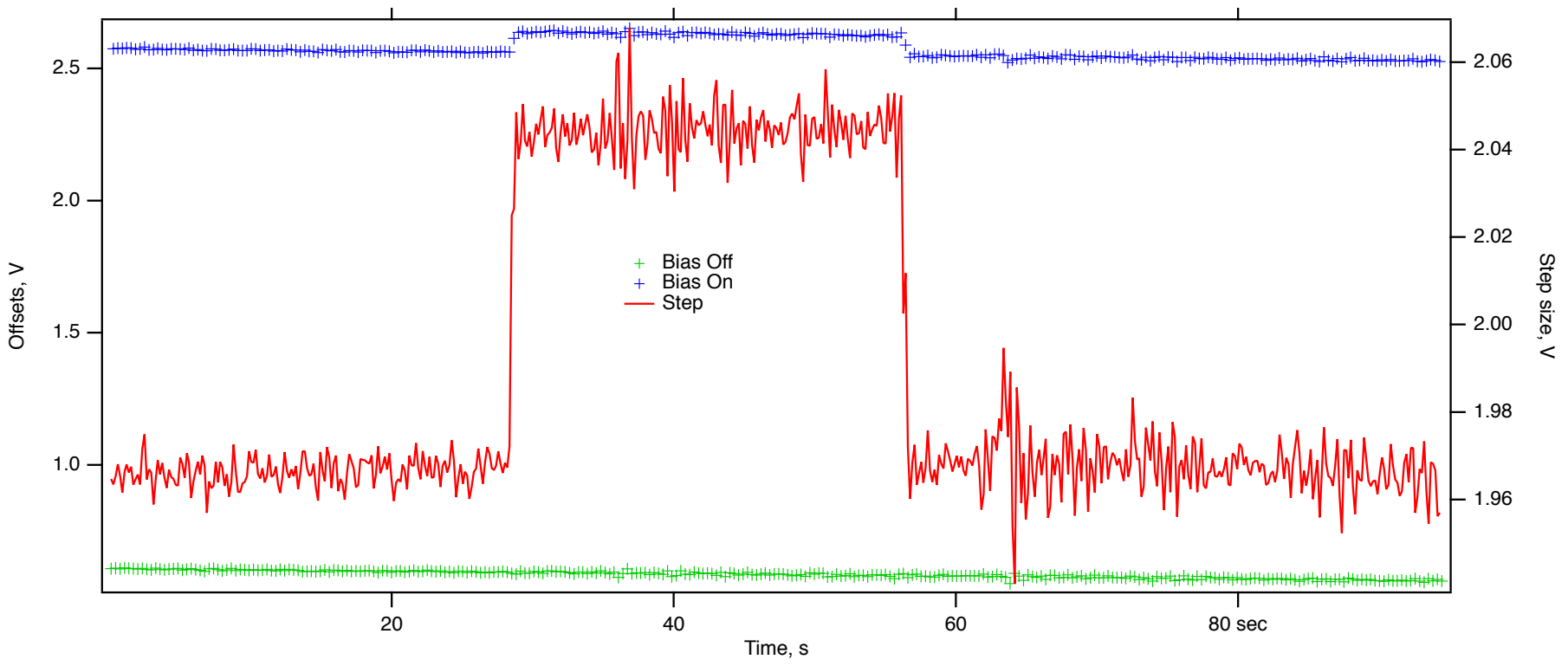

Figure 5. The equilibrium position following each step.

A massive object ( a medium size car ) was brought to within about 3 m of the sensor at t = 28 s and subsequently removed at t = 57 s, and this produced the change observed in the red line. The frequencies obtained do not correspond to any of the known mechanical modes of the ribbon. The Fourier transform of the entire data set is shown in Fig. 6

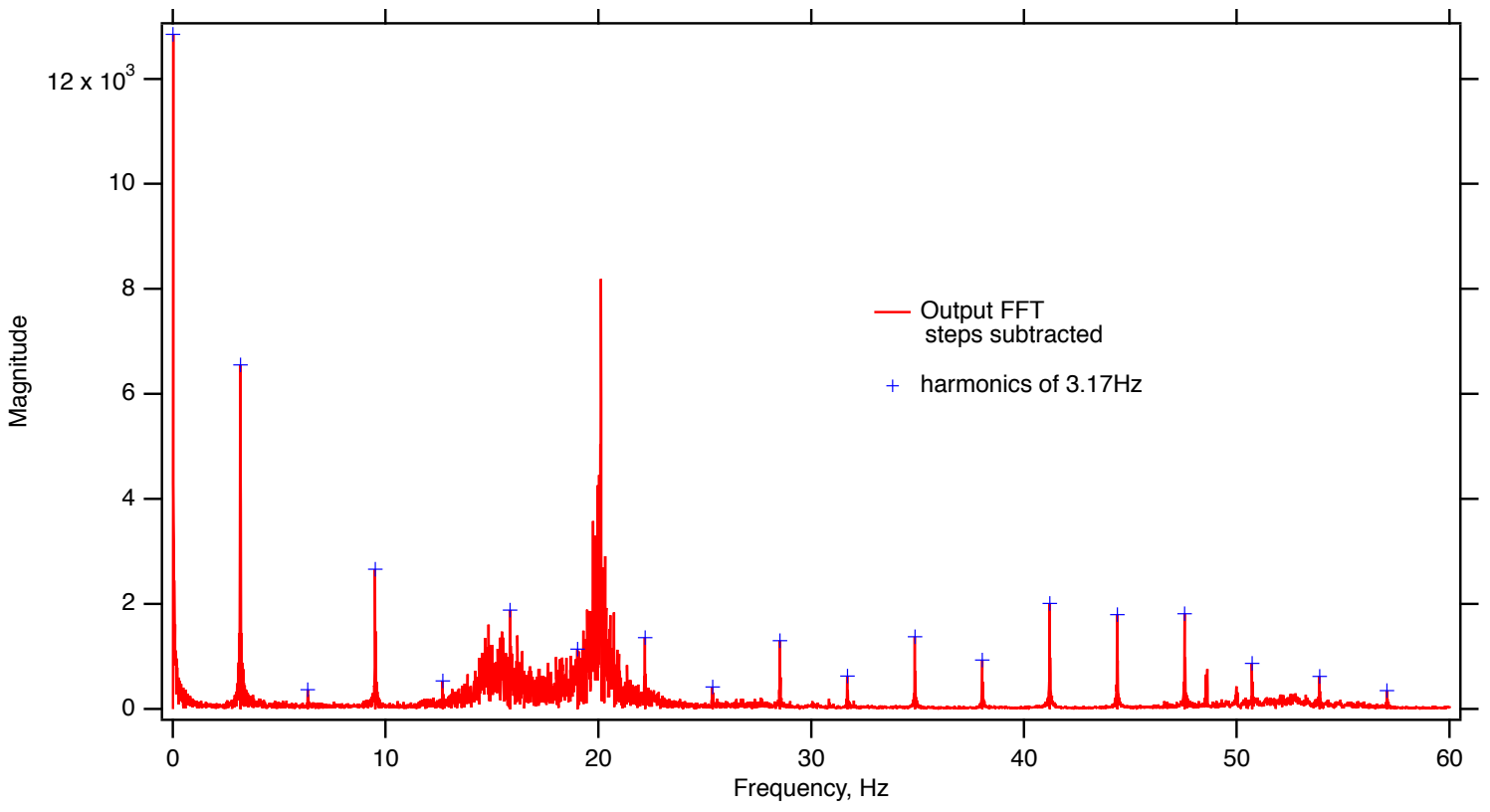

Figure 6. Fourier transform of the data set

The fundamental frequency of the modulation has been suppressed to enhance the dynamic range of the graph but the fundamental and its harmonics, which are depicted by the blue crosses are quite evident. The 45 Hz signal found from the fitting is however not evident, because it is not phase coherent over the entire data set.

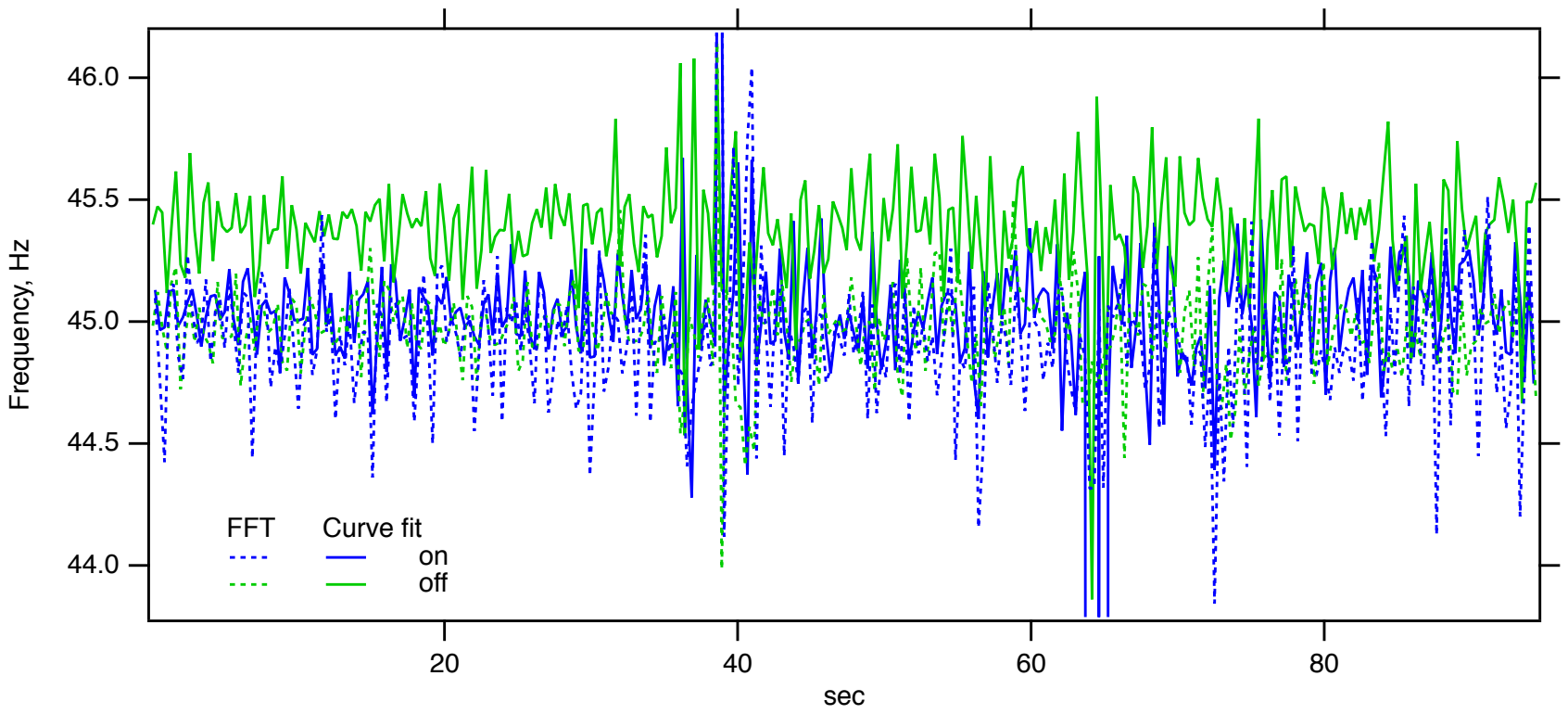

Figure 7. Time series of the frequencies obtained from the fitThe dashed lines show the corresponding frequencies obtained from the Fourier transform, and while these two different measurements are in agreement for the on state, there is a systematic difference for the off state, which is not explained.

The frequencies obtained by fitting are plotted as a time sequence in Fig. 7, where the colour coding is similar to that in Fig. 5. The solid lines show the results of the fit, with green representing the modulation on the off state and blue the on state.

Another view of this is shown in figure 8.

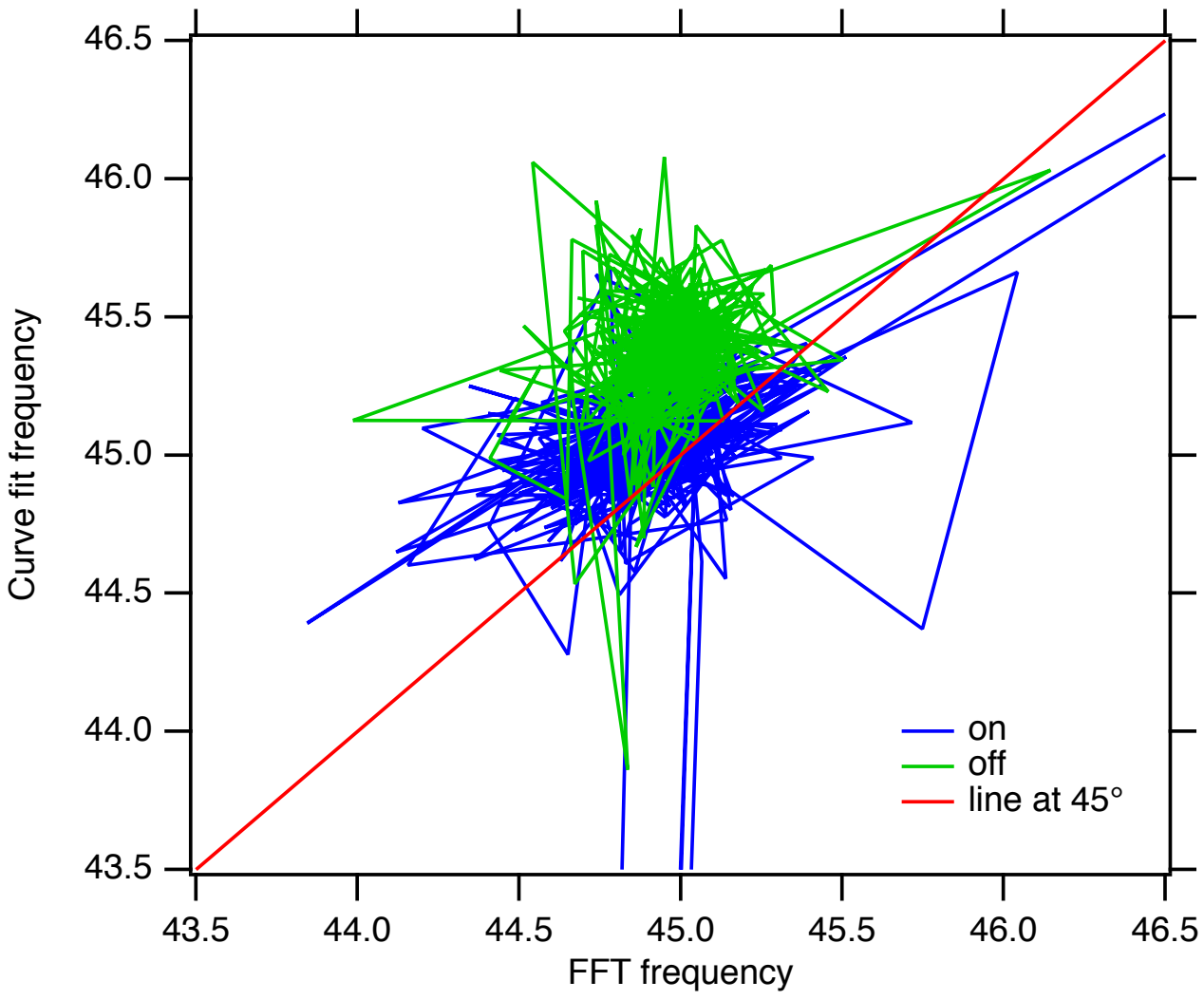

Figure 8. Parametric plot of the frequencies measured by the curve fit and Fourier transform.

If the two methods for obtaining the frequencies are in agreement, the "blobs" would be centred along the red line, plotted at 45°. The reason for this discrepancy is not understood. It might be related to the coarse quantisation of the frequency axis of the Fourier transform, although the position of the peak is obtained by fitting rather than simply taking the frequency at which the peak value occurs.

There are some evident outliers in the frequency plots, a few of which extend beyond the range of the graph. A tool was created to visually examine each step in detail to determine whether the results seem reliable. The two controls at the top of the graph make this convenient, because they permit the graph to be zoomed in to any individual step or optionally some number of consecutive steps. The variable $N$ determines the index of the first step shown and $\Delta N$ determines the number of steps shown. A few typical outliers are shown in Fig. 9. The reason for this behavior is not explained, but it is clear that for these steps, a different mixture of the normal modes was excited.

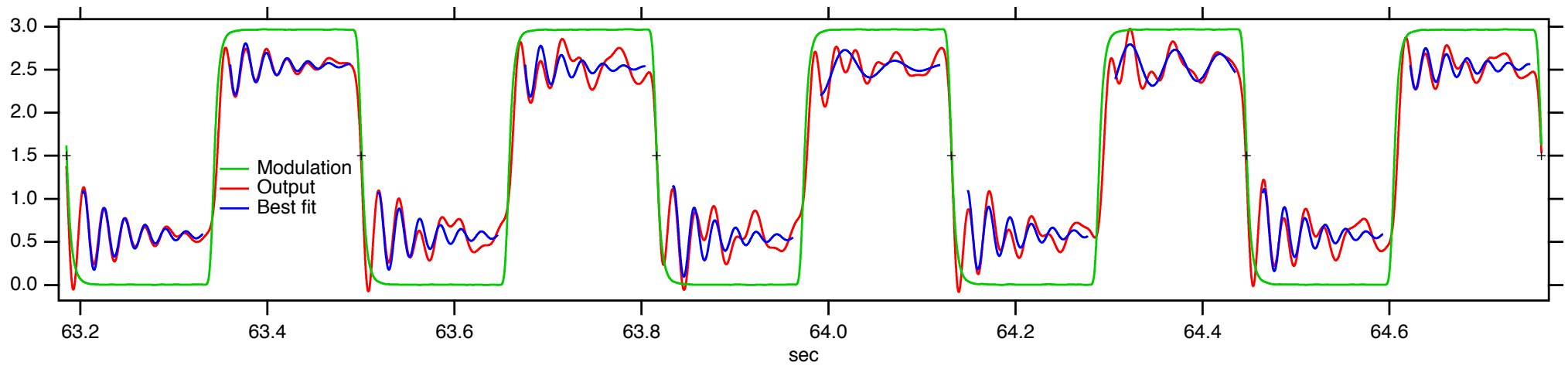

Figure 9. A few outliers

## The Field Data

The field data shows much greater excitation of the mechanical modes. Many modes are excited simultaneously, making it meaningless to fit a single decaying sinusoid. In these circumstances, the most reliable method of determining the equilibrium position is to perform a simple average instead of a fit as depicted in Fig. 10 below.

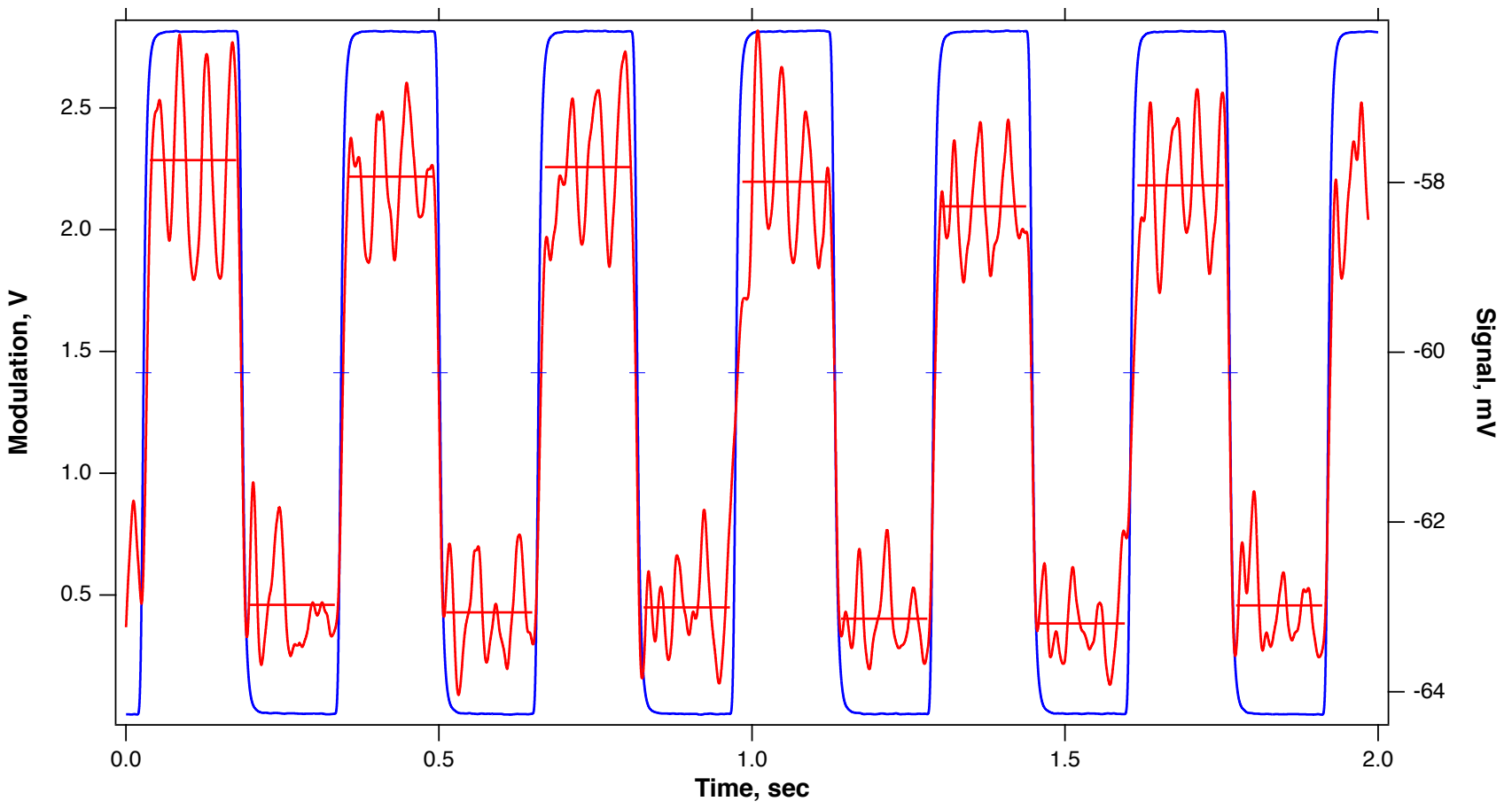

Figure 10. The modulation signal (blue) and output signal (red)

In all cases the field data consists of a large number of individual data files (in .csv format), one for each station and a special signal processing GUI, based on either professional Igor Pro 9 or MatLab software environment, were developed to view the all together in the same graph. The inclinometer data were included in the processing algorithm as shown in Fig. 11 below.

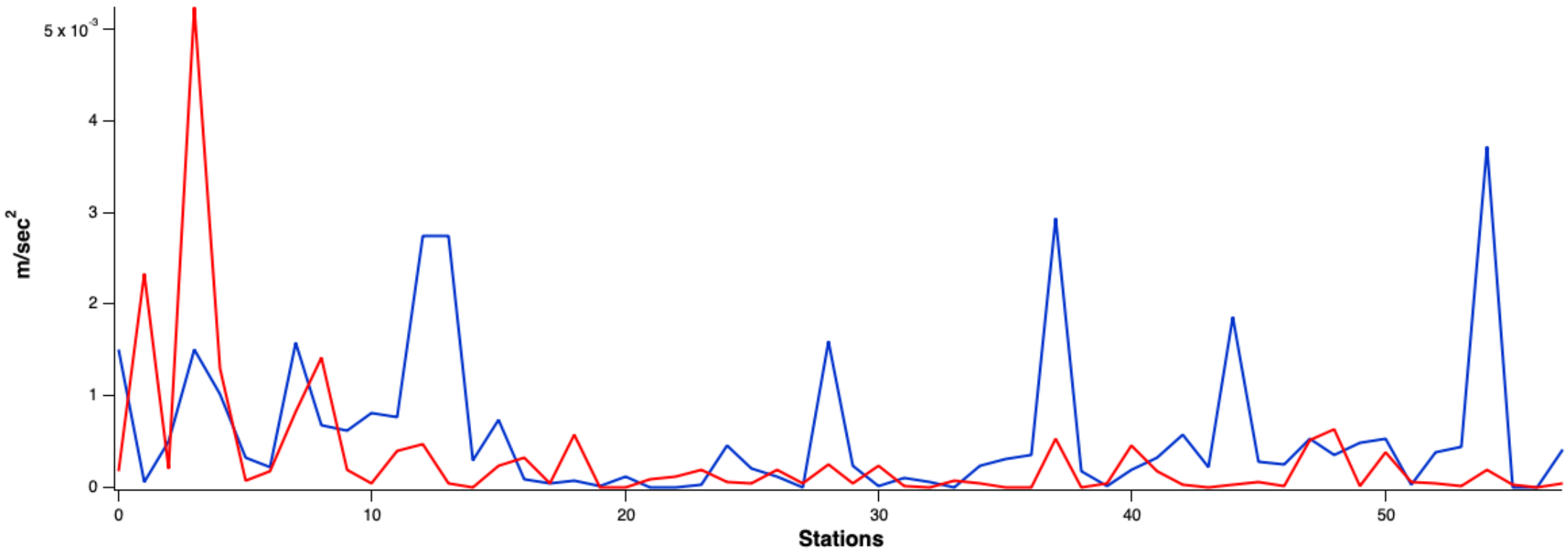

Figure 11. The inclinometer forward data (red) and return reversed data (blue)